\documentclass[12pt,preprint]{aastex}
\def    \mc {\mbox{ cos }}

\def    \mn {\bf N }

\begin{document}
\shorttitle{Subsonic Alignment}
\shortauthors{Lazarian \& Hoang}
\title{Subsonic Mechanical Alignment of Irregular Grains}

\author{A. Lazarian \& Thiem Hoang}
\affil{Dept. of Astronomy, University of Wisconsin,
   Madison, WI53706; lazarian; hoang@astro.wisc.edu}

\begin{abstract}
We show that grains can be efficiently aligned by interacting with a
subsonic gaseous flow. The alignment arises from grains having irregularities that 
scatter atoms with different efficiency in the right and left directions. The grains
tend to align with long axes perpendicular to magnetic field, which corresponds to 
Davis-Greenstein predictions. Choosing conservative estimates, scattering efficiency of impinging atoms and conservative ``degree of helicity'', the alignment of helical grains is much more efficient than the 
Gold-type alignment processes.
\end{abstract}

\section{Introduction}
One of the most convenient ways to trace magnetic field is related to emission
and absorption of polarized radiation by aligned dust. Therefore aligned dust is widely used for this
purpose both in diffuse interstellar medium and molecular gas. Moreover, there 
is evidence of aligned dust around young stellar objects and evolved stars as well as
 astrophysical environments. At the same time, the processes that aligned dust and their relation to magnetic field are still the subject of debates and
require further studies (see Lazarian 2007 for 
a review). While radiative torques (Dolginov \& Mytrophanov 1976, Draine \&
Weingartner 1996, 1997, Weingartner \& Draine 2003, Cho \& Lazarian 2005,
Lazarian \& Hoang 2007, Hoang \& Lazarian 2007) are currently seen as the most 
promising candidate mechanism, the variety of astrophysical conditions may 
enables other mechanisms to dominate in particular environments.

Mechanical alignment was pioneered by Gold (1951, 1952) and then quantified and
elaborated by other researchers (e.g. Purcell 1969, Purcell  \& Spitzer 1971,
Dolginov \& Mytrophanov 1976, Lazarian 1994, 1997, Roberge et al. 1995). While 
the original mechanism could deal with thermally rotating grains only, two 
modifications of the mechanism introduced in Lazarian (1995) and elaborated 
later in Lazarian \& Efroimsky (1996), Lazarian, Efroimsky \& Ozik (1997), Efroimsky (1999) enabled the alignment of grains rotating at much higher rates. The latter were 
introduced to the field by Purcell (1979) (see also Lazarian \& Draine 1999ab,
where the limitations on the size for suprathermally rotating grains are 
discussed).
      
The main shortcoming of the mechanical alignment processes was that they 
required supersonic gas-dust drift to get any appreciable degree of alignment (see Purcell 1969). Although later studies indicated that such drifts can be produced by 
ambipolar diffusion (Roberge \& Hanany 1990, Roberge et al. 1995) or interactions of charged grains with MHD turbulence (Lazarian 1994, Lazarian \& Yan 2002, Yan \& Lazarian 2002, Yan, Lazarian \& Draine 2004), the degree of alignment that is achievable for the
Mach number drifts of the order of unity is insufficient to explain observations 
(see estimates in Lazarian 1997). 

The possibility of subsonic flows to mechanically align grains was mentioned in
passing in Lazarian (2007) and 
Lazarian \& Hoang (2007, henceforth LH07), but no relevant calculations were provided there. 
This possibility is related to irregular grains demonstrating {\it helicity}, i.e.
ability of spin up in a {\it regular way} while interacting with a flow of 
particles. 
\section{Toy model of a helical grain}
In LH07 radiative torques that arise from the interaction of photons with
a grain were considered. Similar torques, however,
 should emerge when atoms bombard the surface
of an irregular grain. 

In Figure~1 we provide a simple model\footnote{Historically, simple models played an important
role for developing models of mechanical alignment. The classical papers by
Gold (1951, 1952) demonstrated the ability of grains to align using a thin
stick as a model of a grain.} of a helical grain. The grain consists of an 
oblate spheroid or an ellipsoid with a mirror attached to it at an angle $\pi/4$. The difference with LH07
is that we assume that the mirror is reflective to atoms rather than photons.

For our purposes in the paper, the actual properties of a
spheroid or an ellipsoid do not matter. Those geometric shapes are too symmetric and
do not produce torques that can spin up a grain in a regular way, i.e. with angular
velocity growing in proportion to time. The stochastic, i.e. with angular velocity growing as a square root of the time, spin up is the essence of the Gold-type alignment 
processes and it is ignored here, where we deal with subsonic flows. For the sake of
simplicity, we assume that the Larmor precession of the grain in the external
magnetic field is much faster than precession arising from the interaction of the ellipsoidal body with the gaseous flow. As a result, following arguments similar to those
in LH07, one can disregard the effects of the gaseous flow on the ellipsoidal body 
altogether. Thus, the only torques to consider are those arising from the mirror.

As in LH07, the the major role of the ellipsoidal body is to provide a steady rotation
about its axis corresponding to the maximal moment of inertia. This ensures that the 
model grain is subject to regular torques while it interacts with the gaseous flow.

The difference between the calculations in LH07 and those
in the current paper stems from
the fact that, while in LH07 the photons are coming as a beam from a single direction,
for subsonic flow that we deal with, atoms hit the mirror from all directions characterized by the flux of atoms given by equation (\ref{ap2}). This 
induces averaging of the torques that we implement below.
\section{Torques on a helical grain}
\begin{figure}
%\plotone{}
\includegraphics[width=0.4\textwidth]{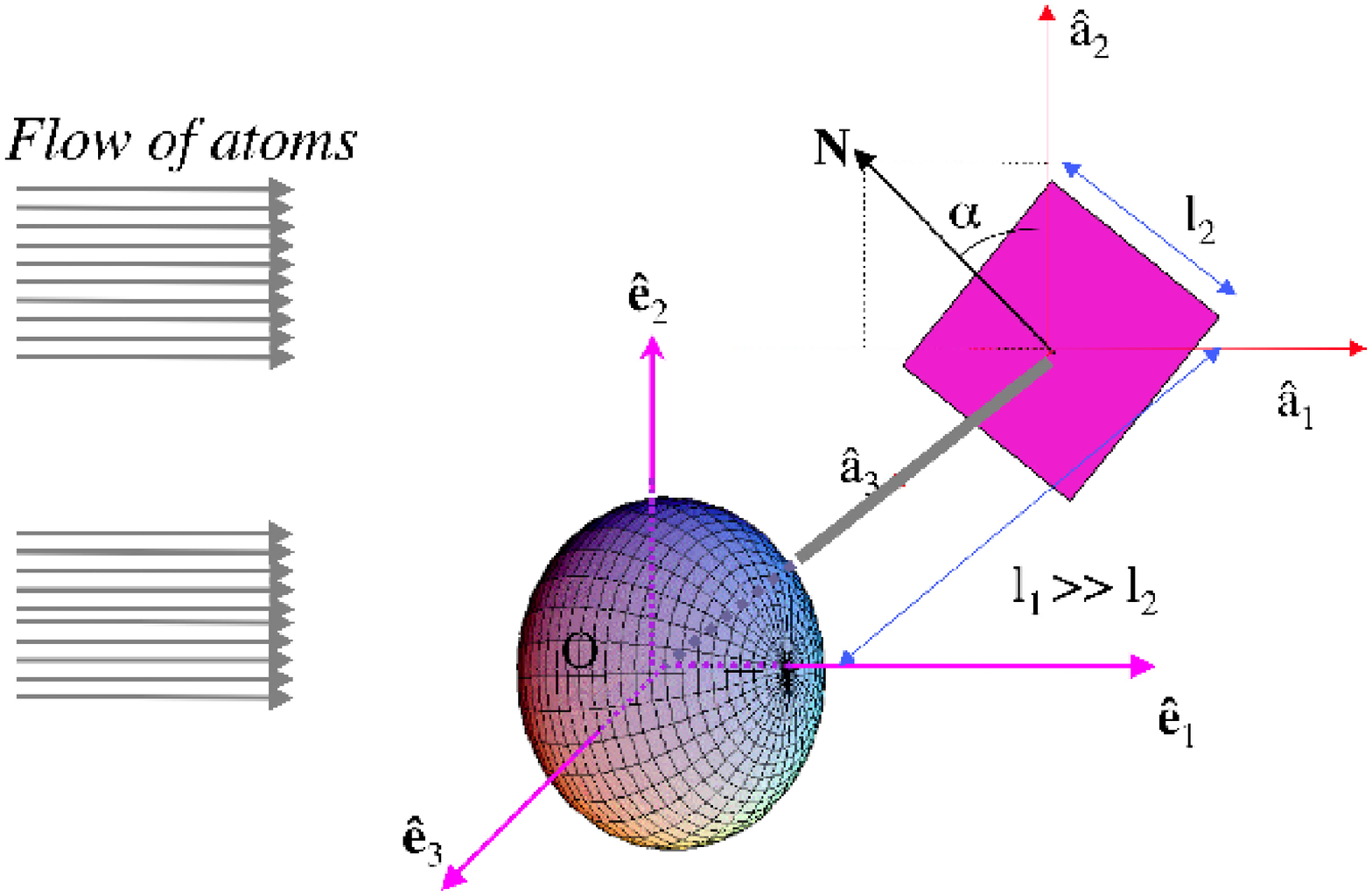}
\caption{A toy model of helical grain consisting of a reflecting spheroid and a mirror attached to it end in gas flows. The distance between the mirror and the spheroid, $l_{1}$ is assumed to be much larger the the mirror size $l_{2}$.}
\label{f0}
\end{figure}

Consider a helical grain (see Fig. \ref{f0}) drifting across a gas chamber with a velocity ${\bf v}_{d}$ along the ${\bf e}_{1}$ axis. The velocity of an atom with respect to the grain is ${\bf v}-{\bf v}_{d}$ where ${\bf v}$ is the thermal velocity of the atom.
The torque from the perfect reflecting of atoms on a surface area $A$ of the mirror is given by
\begin{equation}
d{\bf \Gamma}=[{\bf r}\times \Delta {\bf P}] A f({\bf s}-{\bf s}_{d}) d^{3} s,\label{ap1}
\end{equation}
where ${\bf s}={\bf v}/v_{th}$, vectors ${\bf r}=l_{1} \hat{r}$ is the radius vector directed from the spheroid to the mirror,  $s_{d}=V_{d}/v_{th}$ and $A$ is the surface area of the mirror, and 
\begin{equation}
f({\bf s}-{\bf s}_{d})=n v_{th}({\bf s}-{\bf s}_{d}).{\bf N} e^{-s^{2}},\mbox { ~for~} ({\bf s}-{\bf s}_{d}).{\bf N}<0,  \label{ap2} 
\end{equation}
 is the flux of incoming atoms that can collide with the grain.
Angular momentum element $\Delta {\bf P}$ is antiparallel to the normal vector and given by
\begin{equation}
\Delta {\bf P}=-2 m_{H} {\bf N}|{\bf v}-{\bf v}_{d}|\mc\gamma,\label{ap3}
\end{equation}
where $\mc\gamma=-({\bf v}-{\bf v}_{d}).{\bf N}/|{\bf v}-{\bf v}_{d}|$. Here $\mn$ are radius and normal vector which are functions of angles $\Theta, \beta$ describing the orientation of the grain in the lab system defined by $\hat{\bf e}_{1}, \hat{\bf e}_{2}$ and $\hat{\bf e}_{3}$ (see Fig. \ref{f0}), as given in LH07.\footnote{{\bf N} is given by equation (B6) in LH07, where $n_{1}=-\mbox{sin }\alpha$ and $n_{2}=\mc\alpha$ are components of ${\bf N}$ in the grain coordinate system.}  Substituting equations (\ref{ap2}) and (\ref{ap3}) into equation (\ref{ap1}) we get

\begin{equation}
d{\bf \Gamma}=2 n m_{H} v_{th}^{2} [({\bf r}\times {\bf N}] (({\bf s}-{\bf s}_{d}).{\bf N})^{2} A e^{-s^{2}} d^{3}s,\label{ap4}
\end{equation}
Taking into account the contribution from the reflection on the other surface,  total torque becomes then
\begin{equation}\small
d{\bf \Gamma}=2 n m_{H} v_{th}^{2} [{\bf r}\times {\bf N}]\int_{({\bf s}-{\bf s}_{d}).{\bf N}<0} (({\bf s}-{\bf s}_{d}).{\bf N})^{2} A e^{-s^{2}} d^{3}s\nonumber\\
-2 n m_{H} v_{th}^{2} [{\bf r}\times {\bf N}]\int_{({\bf s}-{\bf s}_{d}).{\bf N}>0} (({\bf s}-{\bf s}_{d}).{\bf N})^{2} A e^{-s^{2}} d^{3}s
,\label{ap5}
\end{equation}
Integral (\ref{ap5}) can be analytically evaluated and the resulting torques are given by 
\begin{equation}
{\bf \Gamma}=\frac{m_{H} n v_{th}^{2} A}{2}[{\bf r}\times {\bf N}]\lbrace |N_{1}| N_{1}(2 s_{d} e^{-s_{d}^{2}}+\sqrt{\pi} erf(s_{d})\nonumber\\
+2 s_{d}^{2} \sqrt{\pi} erf(s_{d}))+\frac{N_{1}}{|N_{1}|}(N_{2}^{2}+N_{3}^{2})erf(s_{d})\rbrace,\label{eq8}
\end{equation}
where $N_{1}, N_{2}$ and $N_{3}$ are components of ${\bf N}$ in the lab coordinate system.

Equation (\ref{eq8}) can be written as
\begin{equation}
{\bf \Gamma}=\frac{m_{H} n v_{th}^{2} A l_{1}}{2}{\bf Q}_{\Gamma},\label{eq9}
\end{equation}
where ${\bf Q}_{\Gamma}$ is the vector torque efficiency consisting of components $Q_{e1}, Q_{e2}$ and $Q_{e3}$ on axes $\hat{\bf e}_{1},\hat{\bf e}_{2}$ and $\hat{\bf e}_{3}$, respectively.

\begin{figure}
\includegraphics[width=0.4\textwidth]{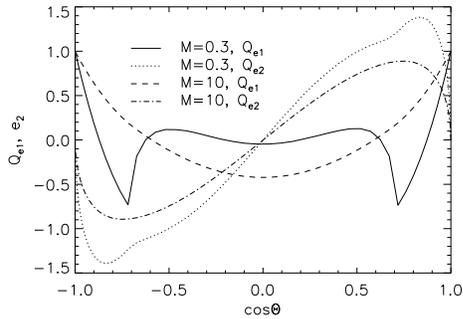}
\caption{The normalized torques $Q_{e1}(\Theta) $ and $Q_{e2}(\Theta)$ averaged over the rotation are
shown for different Mach numbers $M$ of the relative grain-gas flow. For large $M=10$
the torque $Q_{e1}(\Theta) $ is symmetric and $Q_{e2}(\Theta) $ has three zeros at $\Theta=0, \pi/2$ and $\pi$. For subsonic flows $M=0.3$, the torques are modified due to the effect of thermal velocity, but zeros points of $Q_{e2}(\Theta) $ are unchanged.}
\label{f1}
\end{figure}

Two first components for subsonic and supersonic cases are shown in Figure~\ref{f1}. 
The third component produced by the mirror is found to be zero. It is clear that the component $Q_{e3}$ is 
responsible for the precession (LH07). This component is non-zero for a spheroidal body of the grain. In fact, as we discussed above, we disregard this component, as it does not induce either
alignment or spin up.

In Figure \ref{f1} we show torque components for subsonic case $M= 0.3$ and a supersonic case $M=10$. It can be seen that the component $Q_{e2}$ has zero points at $\Theta=0, \pi/2 $ and $\pi$ in both subsonic and supersonic case. The component $Q_{e1}$ is very symmetric in the latter. However, the torques are distorted due to the thermal collisions of gaseous atoms in the former case. The essential properties (i.e. symmetry of $Q_{e1}$ and zero points of $Q_{e2}$) of mechanical torques which are similar to those of radiative torques indicate that mechanical torques can align grains in the same ways as radiative torques do, i.e., grains tend to aligned with long axes perpendicular to magnetic fields. We study this problem in the section below.   
 
\section{Grain alignment with respect to magnetic field {\bf B}}
When the Larmor precession of a grain is larger than the precession induced by mechanical torques (see \S 11.6 in LH07), the alignment occurs with respect to magnetic field (either $\|$ or $\perp$) which defines the axis of alignment.
To study the grain alignment induced by mechanical torques, similar to LH07, we solve the equations of motion for angular momentum ${\bf J}$ in time t:
\begin{equation}
\frac{d{\bf J}}{dt}={\bf \Gamma}-\frac{\bf J}{t_{gas}},\label{eq10}
\end{equation}
where ${\bf \Gamma}$ is given by equation (\ref{eq8}) and $t_{gas}$ is the gaseous damping time. For simplicity, we assume a perfect internal alignment; i.e. the axis of major inertia ${\bf a}_{1}$ is always parallel to ${\bf J}$ (DW97; LH07).
We consider a model grain drifting across magnetic field by an angle $\psi=70^{\circ}$ acted upon by a subsonic flow with $M=0.3$. Representing equation (\ref{eq10}) in the spherical coordinate system defined by $J, \xi, \phi$ where $J$ is the magnitude of ${\bf J}$, $\xi$ is the angle between ${\bf J}$ and ${\bf B}$, and $\phi$ is the Larmor precession angle , we get three equations for $J, \xi$ and $\phi$ (see Sec. 7 in LH07). Averaging over $\phi$ from $0$ to $2\pi$, we obtain only equations for $J $ and $\xi$. We construct trajectory map based on the obtained solution $J(t), \xi(t)$. 

\begin{figure}
\includegraphics[width=0.4\textwidth]{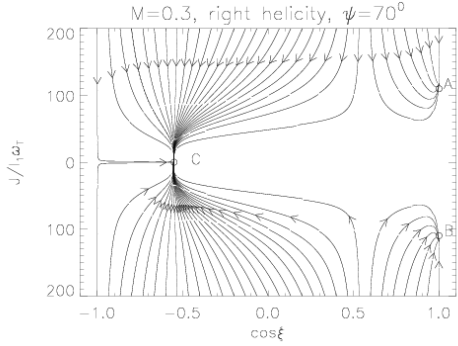}
\caption{Phase trajectory of the grain of $l_{1}=10l_{2}=0.1 \mu m$ and $ED=10^{-3}$, in diffuse gas with $n= 30 cm^{-3}$, temperature
$T=100 K$ corresponding to the March number $M=0.3$ for an angle $\psi$ between ${\bf v}_{d}$ and magnetic field ${\bf B}$ equal $70^{\circ}$ has attractor points A and B of high angular momentum, C with $J=0$ . In the presence of thermal grain wobbling the point C becomes a low-$J$ attractor point with $J\sim J_{therm, dust}$.}
\label{f2}
\end{figure}

Figure \ref{f2} shows the obtained trajectory map with initial condition $J/J_{th}=200$ and the angle $\xi$ generated from a uniform distribution.  We see that grain phase
trajectories that start at different angles $\xi$ converge  on either high-$J$ attractor points A and B corresponding to perfect alignment, or a low-$J$ attractor point C. The latter point formally 
corresponds to $J=0$. However, as it was explained in LH07 and shown by numerical simulations in Hoang \& Lazarian (2007), this is an artifact of our ignoring thermal fluctuations within the grain. If those are taken into account the low-$J$ attractor points correspond to the angular momentum of  the order of the thermal angular momentum corresponding to the temperature of the grain body (see also Weingartner \& Draine 2003, Hoang \& Lazarian 2007).

What should be noted is that the attractor points for $M=0.3$ correspond to the alignment of ${\bf a}_{1}$ parallel to ${\bf B}$; i.e. the grain gets aligned with
long axes perpendicular to magnetic field. This is similar to what the Davis-Greenstein (1951) mechanism of paramagnetic relaxation predicts, but the process we discuss here does not invoke paramagnetic relaxation. 

Gaseous bombardment randomizes grain phase trajectories. However, rather than
tracing phase trajectories, it is practically convenient to have a criterion for the
efficiency of grain alignment. The maximal rotational velocity of grains induced by
torques can be used for this purpose. A study in Hoang \& Lazarian (2007) which solves
the Lagevin equation in order to account for gaseous bombardment shows that
this velocity  which us a function of $\psi$ should be approximately 3 times larger than the thermal rotational 
velocity of the grain.

\begin{figure}
\includegraphics[width=0.4\textwidth]{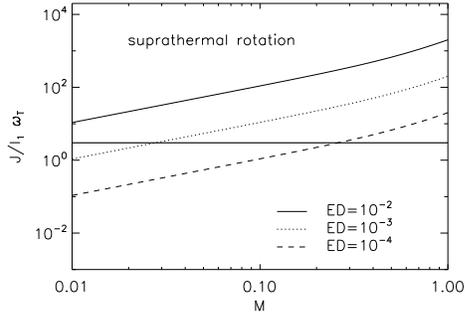}
  \caption{Ratio of the angular momentum induced by grain drifting to the thermal value of the model grain as a function of March number for different ``helicity reduction factors'' $D$. The horizontal line corresponds to $J_{max}(\psi)=3I_{1}\omega_{therm, gas}$ which
presents the threshold for grain alignment obtained in Hoang \& Lazarian (2007). `The results are obtained for a gaseous flow moving parallel to the magnetic field. For grains with velocity mostly perpendicular to the magnetic field (see Yan \& Lazarian 2003), the angular momentum $J$ can be further decreased by a factor 10-100}
\label{f3}
\end{figure}

For practical calculations of the rotational velocity we assume that 10\% of the atoms impinging on the grain mirror are reflected. Figure \ref{f3} shows the angular velocity of a grain
as a function of the Mach number of the flow. The horizontal line corresponds to
$J_{max}=3I_{1}\omega_{therm,gas}$ for $\psi=0^{\circ}$.

It is clear from looking at Figure \ref{f3} that for
our model, helical grains get aligned for relatively low velocities of gas-grain 
drift. All the earlier mechanisms of mechanical alignment are inefficient for such
low subsonic velocities.  Indeed, drifting velocities of the order of $0.1$ of sound speed should be quite common in the ISM (see Yan, Lazarian \& Draine 2004), so potentially the mechanism  is widely applicable. However, it requires further studies.

One may wonder whether the condition of perfect reflection is absolutely necessary.
It can be shown that the torques are changed by a factor of unity if impinging
atoms are absorbed on the grain surface, thermalized there, and emitted from the point that they hit the surface. Moreover, for regular torques to
act on a helical grain, it is sufficient to have a correlation of the place that the
atom impinges on the grain surface and is evaporated from it. It is only when
there is no correlation at all, that our model grain does not experience {\it regular}
torques. Therefore we believe that we do not overestimate the torques with our
assumption of 10\% reflection. In general, to characterize the interaction of grain surface with the gaseous flow, one can introduce a ``reflection efficiency factor'' $E$.

A different issue is the degree of helicity of an irregular grain. Naturally, most
of the grains do not have facets, but have numerous irregularities. An idealized model grain (see Fig. 1), for which the damping arising from the gas interactions with the
ellipsoidal body tends to zero, has the maximal possible 
 value of the torques $Q_{opt}(a, M)$ for
a given grain size $a$ and the Mach number $M$ and its
 ``helicity reduction factor'' $D=1$. This factor, in a general case, can be 
characterized by $D=\frac{Q (a, M)}{Q_{opt}(a, M)} $. 
Further numerical studies should clarify
the value of $D$ for actual irregular grains. In Figure \ref{f3} we show results for different $ED$ factors.
\section{Discussion  and Summary}

In our Letter we use the same toy model of a helical grain as in LH07. In LH07 we proved
the validity of the model (which uses a geometric optics approximation) comparing the functional dependencies of model's torques and
those numerically calculated for the actual irregular grains. In comparison, the 
justification of the model for the gaseous bombardment is self-evident. However,
the combination $ED$ presents a combined uncertainty factor, which characterizes both the uncertainties in grain-gas interactions and grain shape. Testing of the former 
requires laboratory studies, while the latter can be established via the numerical 
research.

What is the relative role of traditional mechanisms of mechanical alignment? Those
mechanisms require supersonic motions to be efficient. It is only mechanical alignment
of helical grains that allows alignment for the subsonic motions. Thus we may claim
that mechanical alignment of helical grains constitutes a new class of mechanical 
alignment processes. In fact, the difference between the mechanical alignment of 
helical grains and the Gold mechanism of grain alignment is similar to the difference
between the radiative torque alignment and the Harwit (1970) process. The latter
appeals to stochastic torques arising from absorption of photons from a light beam and
was shown by Purcell \& Spitzer (1971) to be inefficient for most interstellar
situations.

 Similarly, as the radiative torque alignment is more efficient than the
Harwit (1970) alignment, we believe that the mechanical of helical grains is more
efficient than the Gold processes. Therefore, at supersonic velocities we expect the
mechanical alignment of the actual irregular grains to be governed by their helicity,
rather than the degree of their oblateness of prolateness, which is the case for
the Gold mechanisms. The consequence of this is that the mechanical alignment of
grains will happen with long grain axes perpendicular to magnetic field irrespective
on the gas-grain velocities. This is in contrast to the Gold alignment, for which the
change of alignment happens at the Van Vleck angle $\alpha= 54^{\circ}.7$.

In Hoang \& Lazarian (2007), the effects of gaseous bombardment and uncompensated Purcell's torques (Purcell 1979), e.g. H$_2$ torques,  was studied in connection with the alignment of helical grains subjected to
radiative torques. That study shows that, in the case when both high-$J$ and
low-$J$ attractor points coexist, gaseous bombardment transfers grains from low-$J$ to 
high-$J$ thus increasing\footnote{The increase mostly stems from the higher
degree of internal alignment in high-$J$ attractor points compared to the
low-$J$ ones.}, counter-intuitively, the degree of alignment. This is surely
true only if $J_{max}(\psi)$ is sufficiently high, e.g. higher than
$3I_{1}\omega_{therm, gas}$. The Purcell torques, do not change $J$ in the low-$J$ attractor points because grains flip fast at those points causing thermal
trapping, as discussed in Lazarian \& Draine (1999a). They, however, can increase
the values of $J$ at high-$J$ attractor points. For most  situations, this does not
affect the degree of alignment.

Our considerations above deal with ordinary paramagnetic grains. For those, the
influence of paramagnetic torques is mostly negligible for the typical ISM conditions.
The situation changes if grains are superparamagnetic (see Jones \& Spitzer 1967).
In this situation one can expect always to have alignment with high-$J$ attractor
points, thus having degree of alignment close to 100\%. The assumption of the 
presence of superparamagnetic inclusions is an extra assumption, however. 

As we have stressed above, the mechanical alignment we discuss here and the radiative
torque alignment are different incarnations of the alignment of helical grains.
While the radiative torques have attracted a lot of attention recently, the
mechanical alignment of helical grains has only be mentioned in a couple of
publications (Lazarian 1995, 2007, Lazarian et al. 1997, LH07). 
The relative role of the mechanisms depends on the yet uncertain factor $ED$ for
the mechanical processes. The observations tends to be in agreement with the radiative
torque predictions (see Lazarian 2007 and ref. therein). It is encouraging that
both mechanisms predict the alignment with long grain axes perpendicular to magnetic 
field. Therefore one may hope that the subsonic alignment of helical grains can reveal
via polarimetry the magnetic fields in the situations when the radiative torques fail.
The variety of astrophysical circumstances ensures that there are situations when 
grains are aligned by mechanical subsonic flows. Note that aligned grains were 
reported not only for the ISM, but also comets\footnote{The alignment in comets
may happen in respect to the direction of the gaseous flow rather than to 
magnetic field.} and circumstellar regions. They are also
likely to be present in the disks around evolved and young stars.

The goal of this Letter above is to attract the attention of the community to the
possibility of mechanical alignment of helical dust grains. While further studies of
the process are necessary, the following points can be made at this moment:

1. Irregular grains are, in general, are expected to exhibit helicity when they
interact with gaseous flows. 

2. The mechanical alignment of helical grains is efficient even when the flow of gas
is subsonic. This alignment is likely to dominate Gold-type processes.

3. The mechanical alignment of helical grains aligns grains with long axes perpendicular to magnetic field, thus increasing the chances of magnetic field tracing when other 
alignment mechanisms fail.        
    
{\bf Acknowledgments.}
The work was supported by NSF grant AST 0507164 and the NSF Center for 
Magnetic Self-Organization in Laboratory and Astrophysical Plasmas.

\clearpage

\end{document}